\pdfoutput=1
\RequirePackage{ifpdf}
\ifpdf % We are running pdfTeX in pdf mode
\documentclass[pdftex]{sigma}
\else
\documentclass{sigma}
\fi

\numberwithin{equation}{section}

\newtheorem{Theorem}{Theorem}[section]
\newtheorem{Proposition}[Theorem]{Proposition}
 { \theoremstyle{definition}
\newtheorem{Remark}[Theorem]{Remark} }

\begin{document}

\allowdisplaybreaks

\newcommand{\arXivNumber}{1705.00518}

\renewcommand{\thefootnote}{}

\renewcommand{\PaperNumber}{051}

\FirstPageHeading

\ShortArticleName{Self-Dual Systems, their Symmetries and Reductions to the Bogoyavlensky Lattice}

\ArticleName{Self-Dual Systems, their Symmetries\\ and Reductions to the Bogoyavlensky Lattice\footnote{This paper is a~contribution to the Special Issue on Symmetries and Integrability of Dif\/ference Equations. The full collection is available at \href{http://www.emis.de/journals/SIGMA/SIDE12.html}{http://www.emis.de/journals/SIGMA/SIDE12.html}}}

\Author{Allan P.~FORDY~$^\dag$ and Pavlos XENITIDIS~$^\ddag$}

\AuthorNameForHeading{A.P.~Fordy and P.~Xenitidis}

\Address{$^\dag$~School of Mathematics, University of Leeds, Leeds LS2 9JT, UK}
\EmailD{\href{mailto:A.P.Fordy@leeds.ac.uk}{A.P.Fordy@leeds.ac.uk}}

\Address{$^\ddag$~School of Mathematics, Statistics and Actuarial Science, University of Kent,\\
\hphantom{$^\ddag$}~Canterbury CT2 7FS, UK}
\EmailD{\href{mailto:P.Xenitidis@kent.ac.uk}{P.Xenitidis@kent.ac.uk}}

\ArticleDates{Received May 01, 2017, in f\/inal form June 26, 2017; Published online July 06, 2017}

\Abstract{We recently introduced a class of ${\mathbb{Z}}_N$ graded discrete Lax pairs and studied the associated discrete integrable systems (lattice equations). In particular, we introduced a~subclass, which we called ``self-dual''. In this paper we discuss the continuous symmetries of these systems, their reductions and the relation of the latter to the Bogoyavlensky equation.}

\Keywords{discrete integrable system; Lax pair; symmetry; Bogoyavlensky system}

\Classification{37K05; 37K10; 37K35; 39A05}

\renewcommand{\thefootnote}{\arabic{footnote}}
\setcounter{footnote}{0}

\section{Introduction}

We recently introduced a class of ${\mathbb{Z}}_N$ graded discrete Lax pairs and studied the associated discrete integrable systems \cite{f14-3,f17-2}. Many well known examples belong to that scheme for $N=2$, so, for $N\geq 3$, some of our systems may be regarded as generalisations of these.

In this paper we give a short review of our considerations and discuss the general framework for the derivation of continuous f\/lows compatible with our discrete Lax pairs. These derivations lead to dif\/ferential-dif\/ference equations which def\/ine generalised symmetries of our systems \cite{f14-3}. Here we are interested in the particular subclass of self-dual discrete integrable systems, which exist only for $N$ odd \cite{f17-2}, and derive their lowest order generalised symmetries which are of order one. We also derive corresponding master symmetries which allow us to construct inf\/inite hierarchies of symmetries of increasing orders.

These self-dual systems also have the interesting property that they can be {\it reduced} from $N-1$ to $\frac{N-1}{2}$ components, still with an $N\times N$ Lax pair. However not all symmetries of our original systems are compatible with this reduction. From the inf\/inite hierarchies of generalised symmetries only the even indexed ones are reduced to corresponding symmetries of the reduced systems. Thus the lowest order symmetries of our reduced systems are of order two.

Another interesting property of these dif\/ferential-dif\/ference equations is that they can be brought to a polynomial form through a Miura transformation. In the lowest dimensional case ($N=3$) this polynomial equation is directly related to the Bogoyavlensky equation (see (\ref{eq:Bog}) below), whilst the higher dimensional cases can be regarded as multicomponent generalisations of the Bogoyavlensky lattice.

Our paper is organised as follows. Section \ref{sec:ZN-LP} contains a short review of our framework, the fully discrete Lax pairs along with the corresponding systems of dif\/ference equations, and Section~\ref{continuous-defs} discusses continuous f\/lows and symmetries. The following section discusses the self-dual case and the reduction of these systems. It also presents the systems and corresponding reductions for $N=3$, $5$ and $7$. Section~\ref{sec:Miura} presents the Miura transformations for the reduced systems in $N=3$ and $5$, and discusses the general formulation of these transformations for any dimension~$N$.

\section[${\mathbb{Z}}_N$-graded Lax pairs]{$\boldsymbol{{\mathbb{Z}}_N}$-graded Lax pairs} \label{sec:ZN-LP}

We now consider the specif\/ic discrete Lax pairs, which we introduced in \cite{f14-3,f17-2}. Consider a pair of matrix equations of the form
\begin{subequations}\label{eq:dLP-gen}
\begin{gather}
 \Psi_{m+1,n} = L_{m,n} \Psi_{m,n} \equiv \big( U_{m,n} + \lambda \Omega^{\ell_1}\big) \Psi_{m,n}, \label{eq:dLP-gen-L} \\
 \Psi_{m,n+1} = M_{m,n} \Psi_{m,n} \equiv \big( V_{m,n} + \lambda \Omega^{\ell_2}\big) \Psi_{m,n}, \label{eq:dLP-gen-M}
\end{gather}
where
\begin{gather} \label{eq:A-B-entries}
U_{m,n} = \operatorname{diag}\big(u^{(0)}_{m,n},\dots,u^{(N-1)}_{m,n}\big) \Omega^{k_1},\qquad
 V_{m,n} = \operatorname{diag}\big(v^{(0)}_{m,n},\dots,v^{(N-1)}_{m,n}\big) \Omega^{k_2},
\end{gather}
and
\begin{gather*}
(\Omega)_{i,j} = \delta_{j-i,1} + \delta_{i-j,N-1}.
\end{gather*}
\end{subequations}
The matrix $\Omega$ def\/ines a grading and the four matrices of (\ref{eq:dLP-gen}) are said to be of respective le\-vels~$k_i$,~$\ell_i$, with $\ell_i\neq k_i$ (for each~$i$). The Lax pair is characterised by the quadruple $(k_1,\ell_1;k_2,\ell_2)$, which we refer to as {\em the level structure} of the system, and for consistency, we require
\begin{gather*}%\label{eq:dLP-nec-rel}
k_1 + \ell_2 \equiv k_2 + \ell_1 \quad (\bmod N).
\end{gather*}
Since matrices $U$, $V$ and $\Omega$ are independent of $\lambda$, the compatibility condition of (\ref{eq:dLP-gen}),
\begin{gather*} %\label{eq:dLP-gen-cc}
L_{m,n+1} M_{m,n} = M_{m+1,n} L_{m,n},
\end{gather*}
splits into the system
\begin{subequations} \label{eq:dLP-gen-scc}
\begin{gather}
U_{m,n+1} V_{m,n} = V_{m+1,n} U_{m,n} , \label{eq:dLP-gen-scc-1}\\
U_{m,n+1} \Omega^{\ell_2} - \Omega^{\ell_2} U_{m,n} = V_{m+1,n} \Omega^{\ell_1} - \Omega^{\ell_1} V_{m,n}, \label{eq:dLP-gen-scc-2}
\end{gather}
\end{subequations}
which can be written explicitly as
\begin{subequations} \label{eq:dLP-ex-cc}
\begin{gather}
u^{(i)}_{m,n+1} v_{m,n}^{(i+k_1)} = v^{(i)}_{m+1,n} u^{(i+k_2)}_{m,n} , \label{eq:dLP-ex-cc-1}\\
u^{(i)}_{m,n+1} - u_{m,n}^{(i+\ell_2)} = v^{(i)}_{m+1,n} - v^{(i+\ell_1)}_{m,n} , \label{eq:dLP-ex-cc-2}
\end{gather}
\end{subequations}
for $i \in {\mathbb{Z}}_N$.

\subsection{Quotient potentials}\label{sect:coprime-quotient}

Equations (\ref{eq:dLP-ex-cc-1}) hold identically if we set
\begin{gather} \label{eq:dLP-gen-ph-1}
u^{(i)}_{m,n} = \alpha \frac{\phi^{(i)}_{m+1,n}}{\phi^{(i+k_1)}_{m,n}} ,\qquad v^{(i)}_{m,n} = \beta \frac{\phi^{(i)}_{m,n+1}}{\phi^{(i+k_2)}_{m,n}} ,\qquad i \in {\mathbb{Z}}_N,
\end{gather}
after which (\ref{eq:dLP-ex-cc-2}) takes the form
\begin{gather} \label{eq:dLP-gen-sys-1}
\alpha \left(\frac{\phi^{(i)}_{m+1,n+1}}{\phi^{(i+k_1)}_{m,n+1}} - \frac{\phi^{(i+\ell_2)}_{m+1,n}}{\phi^{(i+\ell_2+k_1)}_{m,n}} \right) =
 \beta \left(\frac{\phi^{(i)}_{m+1,n+1}}{\phi^{(i+k_2)}_{m+1,n}} - \frac{\phi^{(i+\ell_1)}_{m,n+1}}{\phi^{(i+\ell_1+k_2)}_{m,n}} \right) ,
 \qquad i \in {\mathbb{Z}}_N,
\end{gather}
def\/ined on a square lattice. These equations can be explicitly solved for the variables on any of the four vertices and, in particular,
\begin{gather} \label{eq:dLP-gen-sys-1-a}
\phi^{(i)}_{m+1,n+1} = \frac{\phi_{m,n+1}^{(i+k_1)} \phi_{m+1,n}^{(i+k_2)}}{\phi_{m,n}^{(i+k_1+\ell_2)}} \left(
\frac{\alpha \phi_{m+1,n}^{(i+\ell_2)}- \beta \phi_{m,n+1}^{(i+\ell_1)}}{\alpha \phi_{m+1,n}^{(i+k_2)}- \beta \phi_{m,n+1}^{(i+k_1)}}
\right) ,\qquad i \in {\mathbb{Z}}_N.
\end{gather}
In this potential form, the Lax pair (\ref{eq:dLP-gen}) can be written
\begin{subequations} \label{eq:LP-ir-g-rat}
\begin{gather}
\Psi_{m+1,n} = \big( \alpha {\boldsymbol{\phi}}_{m+1,n} \Omega^{k_1} {\boldsymbol{\phi}}_{m,n}^{-1} + \lambda \Omega^{\ell_1}\big) \Psi_{m,n},\nonumber\\
\Psi_{m,n+1} = \big( \beta {\boldsymbol{\phi}}_{m,n+1} \Omega^{k_2} {\boldsymbol{\phi}}_{m,n}^{-1} + \lambda \Omega^{\ell_2}\big) \Psi_{m,n}, \label{eq:LP-ir-g-rat-1}
\end{gather}
where
\begin{gather} \label{eq:LP-ir-g-rat-2}
{\boldsymbol{\phi}}_{m,n} := \operatorname{diag}\big(\phi^{(0)}_{m,n},\dots,\phi^{(N-1)}_{m,n}\big) \qquad {\mbox{and}} \qquad \det\left({\boldsymbol{\phi}}_{m,n}\right) = \prod_{i=0}^{N-1}\phi^{(i)}_{m,n}=1.
\end{gather}
\end{subequations}
We can then show that the Lax pair (\ref{eq:LP-ir-g-rat}) is compatible if and only if the system (\ref{eq:dLP-gen-sys-1}) holds.

\section{Dif\/ferential-dif\/ference equations as symmetries}\label{continuous-defs}

Here we brief\/ly outline the construction of {\it continuous} isospectral f\/lows of the Lax equa\-tions~(\ref{eq:dLP-gen}), since these def\/ine continuous symmetries for the systems (\ref{eq:dLP-ex-cc}). The most important formula for us is (\ref{eq:phi-sys-sym}), which gives the explicit form of the symmetries in potential form.

We seek continuous time evolutions of the form
%\begin{subequations}
\begin{gather*}\label{psit}
\partial_{t} \Psi_{m,n} = S_{m,n} \Psi_{m,n},
\end{gather*}
which are compatible with each of the discrete shifts def\/ined by (\ref{eq:dLP-gen}), if
\begin{gather*}
\partial_t L_{m,n} = S_{m+1,n} L_{m,n} - L_{m,n}S_{m,n}, \nonumber\\
\partial_t M_{m,n} = S_{m,n+1} M_{m,n} - M_{m,n}S_{m,n}. %\label{LtMt}
\end{gather*}
%\end{subequations}
Since
\begin{gather*}
\partial_t (L_{m,n+1} M_{m,n} - M_{m+1,n} L_{m,n}) \\
\qquad {}= S_{m+1,n+1} (L_{m,n+1} M_{m,n} - M_{m+1,n} L_{m,n}) - (L_{m,n+1} M_{m,n} - M_{m+1,n} L_{m,n}) S_{m,n},
\end{gather*}
we have compatibility on solutions of the fully discrete system (\ref{eq:dLP-gen-scc}).

If we def\/ine $S_{mn}$ by
\begin{gather}\label{Q=LS}
S_{m,n} = L_{m,n}^{-1}Q_{m,n},\qquad\mbox{where}\quad Q_{m,n}=\operatorname{diag}\big(q^{(0)}_{m,n},q^{(1)}_{m,n},\dots ,q^{(N-1)}_{m,n}\big) \Omega^{k_1},
\end{gather}
then
\begin{gather*}
Q_{m,n}U_{m-1,n} - U_{m,n} \Omega^{-\ell_1} Q_{m,n}\Omega^{\ell_1} = 0, \qquad  \partial_{t} U_{m,n} = \Omega^{-\ell_1} Q_{m+1,n}\Omega^{\ell_1} - Q_{m,n},
\end{gather*}
which are written explicitly as
\begin{subequations}\label{X1}
\begin{gather}
q^{(i)}_{m,n} u^{(i+k_1)}_{m-1,n} = u^{(i)}_{m,n} q^{(i+k_1-\ell_1)}_{m,n}, \label{q-eqs} \\
\partial_t u^{(i)}_{m,n} = q^{(i-\ell_1)}_{m+1,n} - q^{(i)}_{m,n}. \label{eq:gen-eq-sym}
\end{gather}
It is also possible to prove (see~\cite{f14-3}) that
\begin{gather}\label{tracecon1}
 \sum_{i=0}^{N-1} \frac{q_{m,n}^{(i)}}{u_{m,n}^{(i)}} = \frac{1}{\alpha^N},
\end{gather}
\end{subequations}
for an autonomous symmetry.

Equations (\ref{q-eqs}) and (\ref{tracecon1}), fully determine the functions $q^{(i)}_{m,n}$ in terms of ${\bf u}_{m,n}$ and ${\bf u}_{m-1,n}$.

\begin{Remark}[symmetries in $m$- and $n$-directions]
The formula (\ref{Q=LS}) def\/ines a symmetry which (before prolongation) only involves shifts in the $m$-direction. There is an analogous symmetry in the $n$-direction, def\/ined by
\begin{gather}\label{n-sym}
\partial_s \Psi_{m,n} = \big(V_{m,n}+\lambda \Omega^{\ell_2}\big)^{-1} R_{m,n} \Psi_{m,n},\qquad\mbox{with}\quad R_{m,n} = \operatorname{diag}\big(r^{(0)}_{m,n},\dots,r^{(N-1)}_{m,n}\big) \Omega^{k_2}.\!\!\!
\end{gather}
\end{Remark}
\subsubsection*{Master symmetry and the hierarchy of symmetries}

The vector f\/ield $X^M$, def\/ined by the evolution
\begin{gather*}%\label{master}
\partial_{\tau} u^{(i)}_{m,n} = (m+1) q^{(i-\ell)}_{m+1,n} - m q^{(i)}_{m,n},\qquad \partial_\tau \alpha = 1/\big(N \alpha^{N-1}\big),
\end{gather*}
with $q^{(i)}$ being the solution of (\ref{q-eqs}) and (\ref{tracecon1}), is a {\em master symmetry} of $X^1$, satisfying
\begin{gather*}
\big[\big[X^M,X^1\big],X^1\big]=0, \qquad\mbox{with}\quad \big[X^M,X^1\big]\neq 0.
\end{gather*}
We then def\/ine $X^k$ recursively by $X^{k+1}=[X^M,X^k]$. We have

\begin{Proposition}
Given the sequence of vector fields $X^k$, defined above, we suppose that, for some $\ell\ge 2$, $\{X^1,\dots ,X^\ell\}$ pairwise commute. Then $[X^i,X^{\ell+1}]=0$, for $1\leq i\leq \ell-1$.
\end{Proposition}
This follows from an application of the Jacobi identity.

\begin{Remark}
We {\it cannot} deduce that $[[X^M,X^\ell],X^\ell]=0$ by using the Jacobi identity. Since we are {\it given} this equality for $\ell=2$, we {\it can} deduce that $[X^1,X^3]=0$ (see the discussion around Theorem~19 of~\cite{Y}). Nevertheless it {\it is} possible to check this by hand for low values of $\ell$, for all the examples given in this paper.
\end{Remark}

\subsection{Symmetries in potentials variables}

If we write (\ref{X1}) and the corresponding $n$-direction symmetry in the potential variables (\ref{eq:dLP-gen-ph-1}), we obtain
\begin{subequations}\label{eq:phi-sys-sym}
\begin{gather}
\partial_t \phi^{(i)}_{m,n} = \alpha^{-1} q^{(i-\ell_1)}_{m,n} \phi_{m-1,n}^{(i+k_1)} -\frac{\phi^{(i)}_{m,n}}{N\alpha^{N}} , \label{eq:phi-sys-sym-1} \\
\partial_s \phi^{(i)}_{m,n} = \beta^{-1} q^{(i-\ell_2)}_{m,n} \phi_{m,n-1}^{(i+k_2)} -\frac{\phi^{(i)}_{m,n}}{N\beta^{N}}. \label{eq:phi-sys-sym-2}
\end{gather}
\end{subequations}

\begin{Remark}
The symmetry (\ref{eq:phi-sys-sym-1}) is a combination of the ``generalised symmetry'' (\ref{X1}) and a simple scaling symmetry, with coef\/f\/icient chosen so that the vector f\/ield is {\em tangent} to the level surfaces $\prod\limits_{i=0}^{N-1} \phi^{(i)}_{m,n}=\mbox{const}$, so this symmetry survives the reduction to $N-1$ components, which we always make in our examples. The symmetry~(\ref{eq:phi-sys-sym-2}) is similarly related to (\ref{n-sym}) and also survives the reduction to $N-1$ components.
\end{Remark}

The {\em master symmetries} are similarly adjusted, to give
\begin{subequations}\label{eq:phi-sys-msym}
\begin{gather}
\partial_{\tau} \phi^{(i)}_{m,n} = m \alpha^{-1} q^{(i-\ell_1)}_{m,n} \phi_{m-1,n}^{(i+k_1)}
 -\frac{m \phi^{(i)}_{m,n}}{N\alpha^{N}} , \label{dtauphim} \\
\partial_{\sigma} \phi^{(i)}_{m,n} = n \beta^{-1} q^{(i-\ell_2)}_{m,n} \phi_{m,n-1}^{(i+k_2)}
 -\frac{n \phi^{(i)}_{m,n}}{N\beta^{N}}, \label{dsigmaphim}
\end{gather}
where $\partial_\tau \alpha = 1/\big(N \alpha^{N-1}\big)$ and $\partial_\sigma \beta = 1/\big(N \beta^{N-1}\big)$.
\end{subequations}

\section{The self-dual case}\label{sect:selfdual}

In \cite{f17-2} we give a number of equivalence relations for our general discrete system. For the case with $(k_2,\ell_2)=(k_1,\ell_1)=(k,\ell)$ the mapping
\begin{subequations}\label{sd}
\begin{gather}\label{sd-kl}
(k,\ell)\mapsto \big(\tilde k,\tilde \ell\big) = (N-\ell,N-k)
\end{gather}
is an involution on the parameters, so we refer to such systems as {\em dual}. The {\em self-dual} case is when $(\tilde k,\tilde \ell)=(k,\ell)$, giving $k+\ell=N$. In particular, we consider the case with
\begin{gather}\label{s-dual}
k+\ell=N,\qquad \ell-k =1 \quad\Rightarrow\quad N=2k+1,
\end{gather}
so we require that $N$ is {\it odd}. In this case, we have that Equations (\ref{eq:dLP-gen-sys-1}) are invariant under the change
\begin{gather}\label{sd-phi}
\big(\phi^{(i)}_{m,n},\alpha,\beta\big) \mapsto \big(\widetilde\phi^{(i)}_{m,n},\widetilde\alpha,\widetilde\beta\big),\qquad\mbox{where}\quad \widetilde\alpha \alpha =1,\quad \widetilde\beta \beta =1,\quad \widetilde{\phi}^{(i)}_{m,n} \phi^{(2k-1-i)}_{m,n} = 1.
\end{gather}
\end{subequations}
The self-dual case admits the reduction $\widetilde{\phi}^{(i)}_{m,n}=\phi^{(i)}_{m,n}$, when $\alpha=-\beta$ ($=1$, without loss of generality), which we write as
\begin{gather*}
\phi^{(i+k)}_{m,n} \phi^{(k-1-i)}_{m,n} = 1,\qquad i=0,\dots ,k-1.
\end{gather*}
The condition $\prod\limits_{i=0}^{N-1} \phi^{(i)}_{m,n} = 1$ then implies $\phi^{(N-1)}_{m,n}=1$. Therefore the matrices $U_{m,n}$ and $V_{m,n}$ are built from $k$ components:
\begin{gather*}
U_{m,n} = \operatorname{diag}\Bigg(\phi^{(0)}_{m+1,n}\phi^{(k-1)}_{m,n},\dots,\phi^{(k-1)}_{m+1,n}\phi^{(0)}_{m,n},
 \frac{1}{\phi^{(k-1)}_{m+1,n}},\frac{1}{\phi^{(0)}_{m,n}\phi^{(k-2)}_{m+1,n}},\dots \\
\hphantom{U_{m,n} = \operatorname{diag}\bigg(}{}\dots ,\frac{1}{\phi^{(k-2)}_{m,n}\phi^{(0)}_{m+1,n}}, \frac{1}{\phi^{(k-1)}_{m,n}}\Bigg)\Omega^k ,
\end{gather*}
with $V_{m,n}$ given by the same formula, but with $(m+1,n)$ replaced by $(m,n+1)$. In this case the system (\ref{eq:dLP-gen-sys-1-a}) reduces to
\begin{gather}\label{self-dual-equn}
\phi^{(i)}_{m+1,n+1} \phi^{(i)}_{m,n} = \frac{1}{\phi^{(k-i-2)}_{m+1,n}\phi^{(k-i-2)}_{m,n+1}} \left(\frac{\phi^{(k-i-2)}_{m+1,n}+\phi^{(k-i-2)}_{m,n+1}}{\phi^{(k-i-1)}_{m+1,n}+\phi^{(k-i-1)}_{m,n+1}}\right),
\quad\mbox{for}\quad i=0,1,\dots , k-1.\!\!\!\!
\end{gather}
\begin{Remark}
This reduction has $\frac{N-1}{2}$ components and is represented by an $N\times N$ Lax pair, but is {\it not} $3D$ consistent.
\end{Remark}

\subsubsection*{Symmetries}

Below we give the explicit forms of the self-dual case for $N=3$, $N=5$ and $N=7$. In each case, we give the lowest order symmetry~$X^1$. However, this symmetry does {\it not} reduce to the case of~(\ref{self-dual-equn}), but the second symmetry, $X^2$, of the hierarchy generated by the master symmetries~(\ref{eq:phi-sys-msym}), is a symmetry of the reduced system.

\subsection[The case $N=3$, with level structure $(1,2;1,2)$]{The case $\boldsymbol{N=3}$, with level structure $\boldsymbol{(1,2;1,2)}$}

After the transformation $\phi^{(0)}_{m,n} \rightarrow 1/\phi^{(0)}_{m,n}$, this system becomes
\begin{subequations} \label{eq:3D-1212}
	\begin{gather}
	\phi^{(0)}_{m+1,n+1} = \frac{\alpha \phi_{m+1,n}^{(1)} - \beta \phi^{(1)}_{m,n+1}}{\alpha \phi_{m+1,n}^{(0)}\phi^{(1)}_{m,n+1} - \beta \phi_{m,n+1}^{(0)}\phi^{(1)}_{m+1,n}} \frac{1}{\phi^{(0)}_{m,n}} ,\\
	\phi^{(1)}_{m+1,n+1} = \frac{\alpha \phi_{m,n+1}^{(0)} - \beta \phi^{(0)}_{m+1,n}}{\alpha \phi_{m+1,n}^{(0)}\phi^{(1)}_{m,n+1} - \beta \phi_{m,n+1}^{(0)}\phi^{(1)}_{m+1,n}} \frac{1}{\phi^{(1)}_{m,n}} .
	\end{gather}
\end{subequations}
System (\ref{eq:3D-1212}) admits two point symmetries generated by
\begin{gather*}
 \begin{cases} \partial_\epsilon \phi^{(0)}_{m,n} = \omega^{n+m} \phi^{(0)}_{m,n}, \\ \partial_\epsilon \phi^{(1)}_{m,n} =  0,\end{cases} \qquad  \begin{cases} \partial_\eta \phi^{(0)}_{m,n} =0,\\ \partial_\eta \phi^{(1)}_{m,n} = \omega^{n+m} \phi^{(1)}_{m,n},\end{cases}\qquad \omega^2+\omega+1=0,
\end{gather*}
and two local generalized symmetries. Here we present the symmetry for the $m$-direction whereas the ones in the $n$-direction follow by changing $\phi^{(i)}_{m+j,n} \rightarrow \phi^{(i)}_{m,n+j}$
%\begin{subequations} %\label{eq:3D-1212-sym-1}
	\begin{gather*}
\partial_{t_1} \phi^{(0)}_{m,n} = \phi^{(0)}_{m,n} \frac{1+\phi^{(0)}_{m+1,n}\phi^{(0)}_{m,n} \phi^{(0)}_{m-1,n} - 2 \phi^{(1)}_{m+1,n} \phi^{(1)}_{m,n} \phi^{(1)}_{m-1,n}}{1+\phi^{(0)}_{m+1,n}\phi^{(0)}_{m,n} \phi^{(0)}_{m-1,n} + \phi^{(1)}_{m+1,n} \phi^{(1)}_{m,n} \phi^{(1)}_{m-1,n}} ,\\
\partial_{t_1} \phi^{(1)}_{m,n} = -\phi^{(1)}_{m,n} \frac{1-2 \phi^{(0)}_{m+1,n}\phi^{(0)}_{m,n} \phi^{(0)}_{m-1,n} + \phi^{(1)}_{m+1,n} \phi^{(1)}_{m,n} \phi^{(1)}_{m-1,n}}{1+\phi^{(0)}_{m+1,n}\phi^{(0)}_{m,n} \phi^{(0)}_{m-1,n} + \phi^{(1)}_{m+1,n} \phi^{(1)}_{m,n} \phi^{(1)}_{m-1,n}} .
	\end{gather*}
We also have the master symmetry (\ref{eq:phi-sys-msym}), which can be written
\begin{gather*}
\partial_\tau \phi^{(0)}_{m,n} = m \partial_{t^1} \phi^{(0)}_{m,n} ,\qquad \partial_\tau \phi^{(1)}_{m,n} = m \partial_{t^1} \phi^{(1)}_{m,n} ,\qquad \partial_\tau \alpha = \alpha,
\end{gather*}
which allows us to construct a hierarchy of symmetries of system (\ref{eq:3D-1212}) in the $m$-direction. For instance, the second symmetry is
\begin{subequations} \label{eq:3D-1212-sym-2}
\begin{gather}
\partial_{t_2} \phi^{(0)}_{m,n} = \frac{\phi^{(0)}_{m,n} \phi^{(1)}_{m+1,n} \phi^{(1)}_{m,n} \phi^{(1)}_{m-1,n}}{{\cal{F}}_{m,n}} ({\cal{S}}_m+1)\left( \frac{({\cal{S}}_m- 1)\big(\phi^{(0)}_{m,n} \phi^{(0)}_{m-1,n} \phi^{(0)}_{m-2,n} \big)}{{\cal{F}}_{m,n} {\cal{F}}_{m-1,n}} \right),\\
\partial_{t_2} \phi^{(1)}_{m,n} = \frac{\phi^{(1)}_{m,n} \phi^{(0)}_{m+1,n} \phi^{(0)}_{m,n} \phi^{(0)}_{m-1,n}}{{\cal{F}}_{m,n}} ({\cal{S}}_m+1)\left( \frac{({\cal{S}}_m- 1)\big(\phi^{(1)}_{m,n} \phi^{(1)}_{m-1,n} \phi^{(1)}_{m-2,n} \big)}{{\cal{F}}_{m,n} {\cal{F}}_{m-1,n}} \right),
\end{gather}
where
\begin{gather}
{\cal{F}}_{m,n} := 1+\phi^{(0)}_{m+1,n}\phi^{(0)}_{m,n} \phi^{(0)}_{m-1,n} + \phi^{(1)}_{m+1,n} \phi^{(1)}_{m,n} \phi^{(1)}_{m-1,n} ,
\end{gather}
and ${\cal{S}}_m$ denotes the shift operator in the $m$-direction.
\end{subequations}

\subsubsection*{The reduced system}

The reduced system (\ref{self-dual-equn}) takes the explicit form (f\/irst introduced in \cite{14-6})
\begin{gather} \label{eq:MX2}
\phi_{m,n} \phi_{m+1,n+1} ( \phi_{m+1,n} + \phi_{m,n+1} ) = 2,
\end{gather}
where
\begin{gather*} %\label{eq:red-3D1212-MX2}
\phi^{(0)}_{m,n} = \phi^{(1)}_{m,n} = \frac{1}{\phi_{m,n}} ,\qquad \beta = - \alpha.
\end{gather*}
With this coordinate, the second symmetry (\ref{eq:3D-1212-sym-2}) takes the form
\begin{gather*}
\partial_{t_2} \phi_{m,n} = \phi_{m,n} \frac{1}{P^{(1)}_{m,n}} \left( \frac{1}{P^{(1)}_{m+1,n}}-\frac{1}{P^{(1)}_{m-1,n}}\right),
\end{gather*}
where
\begin{gather} \label{eq:N3-P-G}
P^{(0)}_{m,n} = \phi_{m+1,n} \phi_{m,n} \phi_{m-1,n} , \qquad P^{(1)}_{m,n} = 2 + P^{(0)}_{m,n},
\end{gather}
f\/irst given in \cite{14-6}. Despite the $t_2$ notation, this is the {\it first} of the hierarchy of symmetries of the reduction~(\ref{eq:MX2}).

\subsection[The case $N=5$, with level structure $(2,3;2,3)$]{The case $\boldsymbol{N=5}$, with level structure $\boldsymbol{(2,3;2,3)}$}

In this case, equations (\ref{eq:dLP-gen-sys-1-a}) take the form
\begin{subequations} \label{eq:N5-sys}
\begin{gather}
\phi ^{(0)}_{m+1,n+1}= \frac{\phi ^{(2)}_{m+1,n} \phi ^{(2)}_{m,n+1}}{\phi^{(0)}_{m,n}} \frac{\alpha \phi ^{(3)}_{m+1,n}-\beta \phi ^{(3)}_{m,n+1}}{\alpha \phi ^{(2)}_{m+1,n}- \beta \phi ^{(2)}_{m,n+1}}, \\
\phi ^{(1)}_{m+1,n+1} = \frac{1}{\phi ^{(1)}_{m,n} \big(\alpha \phi ^{(3)}_{m+1,n}-\beta \phi ^{(3)}_{m,n+1}\big)} \left(\frac{\alpha \phi ^{(3)}_{m,n+1}}{\phi ^{(0)}_{m+1,n}\phi ^{(1)}_{m+1,n} \phi^{(2)}_{m+1,n}} -\frac{\beta \phi ^{(3)}_{m+1,n}}{\phi ^{(0)}_{m,n+1} \phi ^{(1)}_{m,n+1} \phi ^{(2)}_{m,n+1}}\right), \nonumber \\
\phi ^{(2)}_{m+1,n+1} = \frac{\alpha \phi ^{(0)}_{m+1,n}-\beta \phi ^{(0)}_{m,n+1}}{\phi ^{(2)}_{m,n}
\big(\alpha \phi ^{(0)}_{m,n+1} \phi ^{(1)}_{m,n+1} \phi ^{(2)}_{m,n+1} \phi ^{(3)}_{m,n+1}-\beta \phi ^{(0)}_{m+1,n} \phi ^{(1)}_{m+1,n} \phi ^{(2)}_{m+1,n} \phi^{(3)}_{m+1,n}\big)}, \!\!\!\\
\phi ^{(3)}_{m+1,n+1} = \frac{\phi ^{(0)}_{m+1,n} \phi ^{(0)}_{m,n+1}}{ \phi ^{(3)}_{m,n}} \frac{\alpha \phi^{(1)}_{m+1,n} - \beta \phi ^{(1)}_{m,n+1}}{\alpha \phi ^{(0)}_{m+1,n}-\beta\phi ^{(0)}_{m,n+1}}.
\end{gather}
\end{subequations}
Under the transformation (\ref{sd}), the f\/irst and last of these interchange, as do the middle pair.

The lowest order generalised symmetry in the $m$-direction is generated by
\begin{gather*}
\partial_{t_1} \phi^{(i)}_{m,n} = \phi^{(i)}_{m,n} \left(\frac{5 A^{(i)}_{m,n}}{B_{m,n}}-1\right),\qquad i=0,\ldots,3,
\end{gather*}
where, if we denote $F^{(i)}_{m,n} =\phi^{(i)}_{m-1,n} \phi^{(i)}_{m,n} \phi^{(i)}_{m+1,n} $,
\begin{gather*}
 A^{(0)}_{m,n} = F^{(0)}_{m,n} F^{(1)}_{m,n} F^{(2)}_{m,n} \phi^{(1)}_{m,n}\phi^{(2)}_{m,n}\phi^{(3)}_{m,n} ,\qquad
A^{(1)}_{m,n} = F^{(0)}_{m,n} F^{(1)}_{m,n} F^{(2)}_{m,n} F^{(3)}_{m,n} \frac{\phi^{(2)}_{m,n}}{\phi^{(0)}_{n,m}},\\
 A^{(2)}_{m,n} = \phi^{(2)}_{m,n} \phi^{(3)}_{m,n},\qquad A^{(3)}_{m,n} = \phi^{(0)}_{m-1,n} \phi^{(0)}_{m+1,n},\\
 B_{m,n} = \sum_{j=0}^{3} A^{(j)}_{m,n} + F^{(0)}_{m,n} F^{(1)}_{m,n}\phi^{(2)}_{m,n}.
\end{gather*}
The corresponding master symmetry is
\begin{gather*}
\partial_\tau \phi^{(i)}_{m,n} = m \partial_{t_1} \phi^{(i)}_{m,n}, \qquad i= 0,\ldots,3,
\end{gather*}
along \looseness=-1 with $\partial_\tau \alpha = 1$. This is used to construct a hierarchy of symmetries for the system~(\ref{eq:N5-sys}). We omit here the second symmetry as the expressions become cumbersome for the unreduced case.

\subsubsection*{The reduced system}

The reduction (\ref{self-dual-equn}) now has components $\phi^{(0)}_{m,n}$, $\phi^{(1)}_{m,n}$, with $\phi^{(2)}_{m,n}=\frac{1}{\phi^{(1)}_{m,n}}$, $\phi^{(3)}_{m,n}=\frac{1}{\phi^{(0)}_{m,n}}$, $\phi^{(4)}_{m,n}= 1$, and the $2$-component system takes the form
 \begin{subequations}\label{eq:sys-5-red}
\begin{gather}
 \phi^{(0)}_{m+1,n+1} \phi^{(0)}_{m,n} = \frac{1}{\phi^{(0)}_{m+1,n}\phi^{(0)}_{m,n+1}} \left(\frac{\phi^{(0)}_{m+1,n}+\phi^{(0)}_{m,n+1}}{\phi^{(1)}_{m+1,n}+\phi^{(1)}_{m,n+1}}\right),\\[3mm]
 \phi^{(1)}_{m+1,n+1} \phi^{(1)}_{m,n} (\phi^{(0)}_{m+1,n}+\phi^{(0)}_{m,n+1}) = 2.
\end{gather}
\end{subequations}
Only the even indexed generalised symmetries of the system (\ref{eq:N5-sys}) are consistent with this reduction. This means that the lowest order generalised symmetry is
\begin{subequations} \label{eq:sym-self-dual-5a}
\begin{gather}
\partial_{t_2} \phi^{(0)}_{m,n} = \phi^{(0)}_{m,n} \frac{P^{(0)}_{m,n}}{ P^{(2)}_{m,n}} \left( \frac{1}{P^{(2)}_{m+1,n}} - \frac{1}{P^{(2)}_{m-1,n}}\right),\\
\partial_{t_2} \phi^{(1)}_{m,n} = \phi^{(1)}_{m,n} \frac{1}{ P^{(2)}_{m,n}} \left(\frac{1 + P^{(0)}_{m+1,n}}{P^{(2)}_{m+1,n}} - \frac{1+P^{(0)}_{m-1,n}}{P^{(2)}_{m-1,n}} \right),
\end{gather}
\end{subequations}
where
\begin{subequations} \label{eq:N5-P-G}
\begin{gather}
P^{(0)}_{m,n} = \phi^{(0)}_{m-1,n} \phi^{(0)}_{m,n} \phi^{(0)}_{m+1,n} \phi^{(1)}_{m,n}, \\
P^{(1)}_{m,n} = \phi^{(0)}_{m-1,n} \big(\phi^{(0)}_{m,n}\big)^2 \phi^{(0)}_{m+1,n} \phi^{(1)}_{m-1,n} \phi^{(1)}_{m,n} \phi^{(1)}_{m+1,n},\\
P^{(2)}_{m,n} = 2 +2 P^{(0)}_{m,n} + P^{(1)}_{m,n}.\label{eq:G-5}
\end{gather}
\end{subequations}

\subsection[The case $N=7$, with level structure $(3,4;3,4)$]{The case $\boldsymbol{N=7}$, with level structure $\boldsymbol{(3,4;3,4)}$}

\looseness=-1 The fully discrete system (\ref{eq:dLP-gen-sys-1-a}) and its lower order symmetries (\ref{eq:phi-sys-sym}) and master symmetries~(\ref{eq:phi-sys-msym}) can be easily adapted to our choices $N=7$ and $(k,\ell)=(3,4)$. In the same way the correspon\-ding reduced system follows from~(\ref{self-dual-equn}) with $k=3$. Thus we omit all these systems here and present only the lowest order symmetry of the reduced system which takes the following form
\begin{gather}
\partial_{t_2} \phi^{(0)}_{m,n} = \phi^{(0)}_{m,n} \frac{ P^{(1)}_{m,n}}{P^{(3)}_{m,n}} \left(\frac{1}{P^{(3)}_{m+1,n}} - \frac{1}{P^{(3)}_{m-1,n}}\right), \nonumber \\
\partial_{t_2} \phi^{(1)}_{m,n} = \phi^{(1)}_{m,n} \frac{P^{(0)}_{m,n}}{P^{(3)}_{m,n}} \left(\frac{1+ P^{(0)}_{m+1,n}}{P^{(3)}_{m+1,n}} - \frac{1+ P^{(0)}_{m-1,n}}{P^{(3)}_{m-1,n}}\right), \label{eq:N7-red-dd} \\
\partial_{t_2} \phi^{(2)}_{m,n}= \phi^{(2)}_{m,n} \frac{1}{P^{(3)}_{m,n}} \left(\frac{1+ P^{(0)}_{m+1,n}+ P^{(1)}_{m+1,n}}{P^{(3)}_{m+1,n}} - \frac{1+ P^{(0)}_{m-1,n}+ P^{(1)}_{m-1,n}}{P^{(3)}_{m-1,n}}\right), \nonumber
\end{gather}
where
\begin{subequations} \label{eq:N7-P-G}
	\begin{gather}
	P^{(0)}_{m,n} = \phi^{(0)}_{m-1,n} \phi^{(0)}_{m+1,n} \phi^{(1)}_{m,n} \phi^{(2)}_{m,n}, \\
	P^{(1)}_{m,n} = \phi^{(0)}_{m-1,n} \phi^{(0)}_{m,n} \phi^{(0)}_{m+1,n} \phi^{(1)}_{m-1,n} \big(\phi^{(1)}_{m,n}\big)^2 \phi^{(1)}_{m+1,n} \phi^{(2)}_{m,n}, \\
	P^{(2)}_{m,n} =\phi^{(0)}_{m-1,n} \big(\phi^{(0)}_{m,n}\big)^2 \phi^{(0)}_{m+1,n} \phi^{(1)}_{m-1,n} \big(\phi^{(1)}_{m,n}\big)^2 \phi^{(1)}_{m+1,n} \phi^{(2)}_{m-1,n} \phi^{(2)}_{m,n} \phi^{(2)}_{m+1,n},\\
	P^{(3)}_{m,n} = 2 + 2 P^{(0)}_{m,n} + 2 P^{(1)}_{m,n}+ P^{(2)}_{m,n}.
	\end{gather}
\end{subequations}

\section{Miura transformations and relation to Bogoyavlensky lattices} \label{sec:Miura}

In this section we discuss Miura transformations for the reduced systems and their symmetries which bring the latter to polynomial form. In the lowest dimensional case ($N=3$) the polynomial system is directly related to the Bogoyavlensky lattice (see~(\ref{eq:Bog}) below), whereas the higher dimensional ones result in systems which generalise~(\ref{eq:Bog}) to $k$ component systems.

\subsection*{The reduced system in $\boldsymbol{N=3}$}

The Miura transformation \cite{14-6}
\begin{gather*} %\label{eq:Miura-3}
\psi_{m,n} = \frac{P^{(0)}_{m,n}}{P^{(1)}_{m,n}} - 1,
\end{gather*}
where $P^{(0)}_{m,n}$ and $P^{(1)}_{m,n}$ are given in (\ref{eq:N3-P-G}), maps equation (\ref{eq:MX2}) to
\begin{gather} \label{eq:MX2a}
\frac{\psi_{m+1,n+1}+1}{\psi_{m,n}+\psi_{m,n+1}+1} + \frac{\psi_{m+1,n}}{\psi_{m,n+1}} = 0,
\end{gather}
and its symmetry to
\begin{gather}\label{eq:Bog}
\partial_{t_2} \psi_{m,n} = \psi_{m,n} (\psi_{m,n}+1) (\psi_{m+2,n} \psi_{m+1,n} - \psi_{m-1,n} \psi_{m-2,n}),
\end{gather}
which is related to the Bogoyavlensky lattice \cite{B}
\begin{gather*} \partial_{t_2} \chi_{m,n} = \chi_{m,n}(\chi_{m+2,n} + \chi_{m+1,n} - \chi_{m-1,n} - \chi_{m-2,n}),\end{gather*}
through the Miura transformation
\begin{gather*} \chi_{m,n} = \psi_{m+1,n} \psi_{m,n} (\psi_{m-1,n}+1).\end{gather*}

\subsection*{The reduced system in $\boldsymbol{N=5}$}

The Miura transformation
\begin{gather*}
\psi_{m,n}^{(0)} = \frac{2 P^{(0)}_{m,n}}{P^{(2)}_{m,n}},\qquad \psi_{m,n}^{(1)} = \frac{P^{(1)}_{m,n}}{P^{(2)}_{m,n}} - 1,
\end{gather*}
where $P^{(i)}_{m,n}$ are given in (\ref{eq:N5-P-G}), maps system (\ref{eq:sys-5-red}) to
\begin{gather}
\frac{\psi^{(0)}_{m,n+1}\psi^{(0)}_{m+1,n+1}+\psi^{(0)}_{m+1,n} \big(\psi^{(0)}_{m,n}+\psi^{(1)}_{m,n+1}\big)}{\psi^{(1)}_{m,n}+1} = \frac{\psi^{(0)}_{m,n+1}\psi^{(1)}_{m+1,n}-\psi^{(0)}_{m+1,n}\psi^{(1)}_{m,n+1}}{\psi^{(0)}_{m,n+1} + \psi^{(1)}_{m,n+1}}, \nonumber\\
\frac{\psi^{(1)}_{m+1,n+1}+1}{\psi^{(1)}_{m,n}+\psi^{(1)}_{m,n+1}+\psi^{(0)}_{m,n+1}+1} + \frac{\psi^{(0)}_{m+1,n}+\psi^{(1)}_{m+1,n}}{\psi^{(0)}_{m,n+1}+\psi^{(1)}_{m,n+1}} = 0,\label{eq:M-red-sys-5}
\end{gather}
and its symmetry (\ref{eq:sym-self-dual-5a}) to the following system of polynomial equations in which we have suppressed the dependence on the second index~$n$:
	\begin{gather}
	\frac{\partial_{t_2} \psi^{(0)}_m}{\psi^{(0)}_m} = \big(\psi^{(0)}_m+\psi^{(1)}_m\big) \big(\psi^{(0)}_{m+2} \psi^{(0)}_{m+1} - \psi^{(0)}_{m-1} \psi^{(0)}_{m-2} + \psi^{(0)}_{m+1}-\psi^{(0)}_{m-1} + \psi^{(1)}_{m+1}-\psi^{(1)}_{m-1}\big) \nonumber \\
\hphantom{\frac{\partial_{t_2} \psi^{(0)}_m}{\psi^{(0)}_m} =}{} - \big(\psi^{(1)}_m+1\big) \big(\psi^{(1)}_{m+2} \psi^{(1)}_{m+1} - \psi^{(1)}_{m-1} \psi^{(1)}_{m-2}\big) + \big(\psi^{(0)}_m-1\big) \big(\psi^{(1)}_{m+2} \psi^{(0)}_{m+1}-\psi^{(0)}_{m-1} \psi^{(1)}_{m-2}\big), \nonumber\\
	\frac{\partial_{t_2} \psi^{(1)}_m}{\psi^{(1)}_m+1} = \big(\psi^{(0)}_m+\psi^{(1)}_m\big) \big(\psi^{(0)}_{m+2} \psi^{(0)}_{m+1} - \psi^{(0)}_{m-1} \psi^{(0)}_{m-2}\big) - \psi^{(1)}_m \big(\psi^{(1)}_{m+2} \psi^{(1)}_{m+1} - \psi^{(1)}_{m-1} \psi^{(1)}_{m-2}\big) \nonumber \\
 \hphantom{\frac{\partial_{t_2} \psi^{(1)}_m}{\psi^{(1)}_m+1} =}{} + \psi^{(0)}_m \big(\psi^{(1)}_{m+2} \psi^{(0)}_{m+1}-\psi^{(0)}_{m-1} \psi^{(1)}_{m-2}\big).\label{eq:M-sys-5}
	\end{gather}
The above system and its symmetry can be considered as a two-component generalisation of the equation (\ref{eq:MX2a}) and its symmetry (\ref{eq:Bog}) in the following sense. If we set, $\psi^{(0)}_{m,n} = 0$ and $\psi^{(1)}_{m,n} = \psi_{m,n}$ in (\ref{eq:M-red-sys-5}) and~(\ref{eq:M-sys-5}), then they will reduce to equations~(\ref{eq:MX2a}) and~(\ref{eq:Bog}), respectively.

\subsection*{The reduced systems for $\boldsymbol{N>5}$}

It can be easily checked that for each $k$ ($N=2 k+1$), the lowest order symmetry of the reduced system (\ref{self-dual-equn}) involves certain functions $P^{(i)}_{m,n}$, $i=0,\ldots,k$, with
\begin{gather*} P^{(k)}_{m,n} = 2 + 2 \sum_{i=0}^{k-2} P^{(i)}_{m,n} + P^{(k-1)}_{m,n},\end{gather*}
which are given in terms of $\phi^{(i)}_{m,n}$ and their shifts (see relations (\ref{eq:N5-P-G}) and (\ref{eq:N7-P-G})). Then, the Miura transformation
\begin{gather}\label{eq:gen-Miura}
\psi^{(i)}_{m,n} = \frac{2 P^{(i)}_{m,n}}{P^{(k)}_{m,n}}, \qquad i=0,\ldots,k-2, \qquad
\psi^{(k-1)}_{m,n} = \frac{P^{(k-1)}_{m,n}}{P^{(k)}_{m,n}} - 1,
\end{gather}
brings the symmetries of the reduced system to polynomial form. One could derive the polynomial system corresponding to $N=7$ ($k=3$) starting with system (\ref{eq:N7-red-dd}), the functions given in (\ref{eq:N7-P-G}) and using the corresponding Miura transformation~(\ref{eq:gen-Miura}). The system of dif\/ferential-dif\/ference equations is omitted here because of its length but it can be easily checked that if we set $\psi^{(0)}_{m,n}=0$ and then rename the remaining two variables as $\psi^{(i)}_{m,n} \mapsto \psi^{(i-1)}_{m,n}$, then we will end up with system (\ref{eq:M-sys-5}).

This indicates that every $k$ component system is a generalisation of all the lower order ones, and thus of the Bogoyavlensky lattice~(\ref{eq:Bog}). To be more precise, if we consider the case $N=2 k+1$ along with the $k$-component system, set variable $\psi^{(0)}_{m,n}=0$ and then rename the remaining ones as $\psi^{(i)}_{m,n} \mapsto \psi^{(i-1)}_{m,n}$, then the resulting $(k-1)$-component system is the reduced system corresponding to $N = 2 k-1$. Recursively, this means that it also reduces to the $N=3$ system, i.e., equation~(\ref{eq:Bog}). Other systems with similar behaviour have been presented in \cite{BW}.

\subsection*{Acknowledgements}

PX acknowledges support from the EPSRC grant {\it Structure of partial difference equations with continuous symmetries and conservation laws}, EP/I038675/1.
\vspace{-1mm}

\pdfbookmark[1]{References}{ref}
\LastPageEnding

\end{document}